\documentclass[article]{elsarticle}

\usepackage{hyperref}
\usepackage{natbib}

\journal{Journal of \LaTeX\ Templates}

\begin{document}

\begin{frontmatter}

\title{A propagation tool to connect remote-sensing observations with in-situ measurements of heliospheric structures}

\author[aff1,aff2]{A.P. Rouillard}
\cortext[mycorrespondingauthor]{A.P. Rouillard, arouillard@irap.omp.eu}
\author[aff1,aff2]{B. Lavraud}
\author[aff1,aff2]{V. G\'{e}not}
\author[aff1,aff2]{M. Bouchemit}
\author[aff3]{N. Dufourg}
\author[aff1,aff2]{I. Plotnikov}
\author[aff1,aff2]{R.F. Pinto}
\author[aff1,aff2]{E. Sanchez-Diaz}
\author[aff1,aff2]{M. Lavarra}
\author[aff1,aff2]{M. Penou}
\author[aff1,aff2]{C. Jacquey}
\author[aff1,aff2]{N. Andr\'{e}}
\author[aff4]{S. Caussarieu}
\author[aff4]{J.-P. Toniutti}
\author[aff4]{D. Popescu}
\author[aff5]{E. Buchlin}
\author[aff5]{S. Caminade}
\author[aff5,aff6]{P. Alingery}
\author[aff7]{J.A. Davies}
\author[aff8]{D. Odstrcil}
\author[aff9]{L. Mays}
\address[aff1]{Institut de Recherche en Astrophysique et Plan\'{e}tologie, Universit\'{e} de Toulouse III (UPS), France}
\address[aff2]{Centre National de la Recherche Scientifique, UMR 5277, Toulouse, France}
\address[aff3]{Centre National d'Etudes Spatiales, 18 Avenue Edouard Belin, 31400, Toulouse}
\address[aff4]{GFI Informatique, 1 Rond-point du G\'{e}n\'{e}ral Eisenhower, 31100, Toulouse}
\address[aff5]{Institut d'Astrophysique Spatiale, CNRS, Univ. Paris-Sud, Universit\'{e} Paris-Saclay, Bt. 121, 91405 Orsay, France}
\address[aff6]{CesamSeed, 52bis Boulevard Saint Jacques, 75014 Paris, France}
\address[aff7]{Science and Technology Facilities Council, Rutherford Appleton Laboratory, Chilton, UK}
\address[aff8]{George Mason University, 4400 University Dr, Fairfax}
\address[aff9]{NASA Goddard Space Flight Center, 8800 Greenbelt Rd, Greenbelt, USA}


\begin{abstract}
The remoteness of the Sun and the harsh conditions prevailing in the solar corona have so far limited the observational data used in the study of solar physics to remote-sensing observations taken either from the ground or from space. In contrast, the `solar wind laboratory' is directly measured in situ by a fleet of spacecraft measuring the properties of the plasma and magnetic fields at specific points in space. Since 2007, the solar-terrestrial relations observatory (STEREO) has been providing images of the solar wind that flows between the solar corona and spacecraft making in-situ measurements. This has allowed scientists to directly connect processes imaged near the Sun with the subsequent effects measured in the solar wind. This new capability prompted the development of a series of tools and techniques to track heliospheric structures through space. This article presents one of these tools, a web-based interface called the 'Propagation Tool' that offers an integrated research environment to study the evolution of coronal and solar wind structures, such as Coronal Mass Ejections (CMEs), Corotating Interaction Regions (CIRs) and Solar Energetic Particles (SEPs). These structures can be propagated from the Sun outwards to or alternatively inwards from planets and spacecraft situated in the inner and outer heliosphere. In this paper, we present the global architecture of the tool, discuss some of the assumptions made to simulate the evolution of the structures and show how the tool connects to different databases.
\end{abstract}

\begin{keyword}
heliophysics\sep solar imaging\sep CMEs \sep CIRs
\end{keyword}

\end{frontmatter}

\section{A propagation tool for heliospheric research:}
\indent The analysis of remote-sensing observations and in-situ measurements require very different expertise. The physical processes that are remotely sensed must be analysed by first accounting for the mechanisms that generate or scatter the observed electromagnetic radiation. These mechanisms will differ according to the wavelength detected and the height at which the solar atmosphere is observed. In the optically thin corona for instance, each image pixel detects light integrated along the line of sight that passes through an extended region of the corona. At photospheric and chromospheric altitudes the medium is more optically thick. In-situ measurements made further out in the solar wind provide directly the physical parameters of the plasma at a single point but they are exact. Remote-sensing observations are our only source of information of the state of the low corona, solar imagery also provides a large-scale view of the physical processes at play. By combining both remote-sensing observations and in-situ measurements we can gain a more complete picture of the Sun-Earth system over a wide range of spatial and temporal scales. \\

\indent The very different nature of remote-sensing observations and in-situ measurements coupled with the large distances that separate the solar corona with current in-situ measurements, meant that solar and solar wind research evolved separately over many decades, all too often without those involved consulting each other when working on the same problem. The databases used by researchers also evolved independently and little concerted effort was made to link them. Rare attempts have been made to connect some datasets by carrying out ballistic tracing of heliospheric structures between the Sun and planets, or probes. One example of such an undertaking was the first propagation tool developed by the HELIO FP7 project (\url{http://hfe.helio-vo.eu/Helio/}). These first attempts lacked the observational support to validate such ballistic tracing and the user was left uncertain about the validity of the tracing method. The launch of the Heliospheric Imagers (HI) instruments onboard the Solar-Terrestrial Relations Observatory (STEREO) in 2006 has provided imaging of the solar wind with sufficient resolution and cadence to enable plasma structures to be clearly identified and tracked as they propagate through the inner heliosphere (see for example review by Rouillard 2011a). The Propagation Tool developped at the Institute of Research in Astrophysics and Planetology (IRAP), accessible at \url{http://propagationtool.cdpp.eu/}, and presented in this paper exploits, in particular, heliospheric imagery as observational support to connect solar imagery with in-situ measurements.\\

\begin{figure}[ht!]
 \centering
 \includegraphics[width=0.7\textwidth,clip,angle=0]{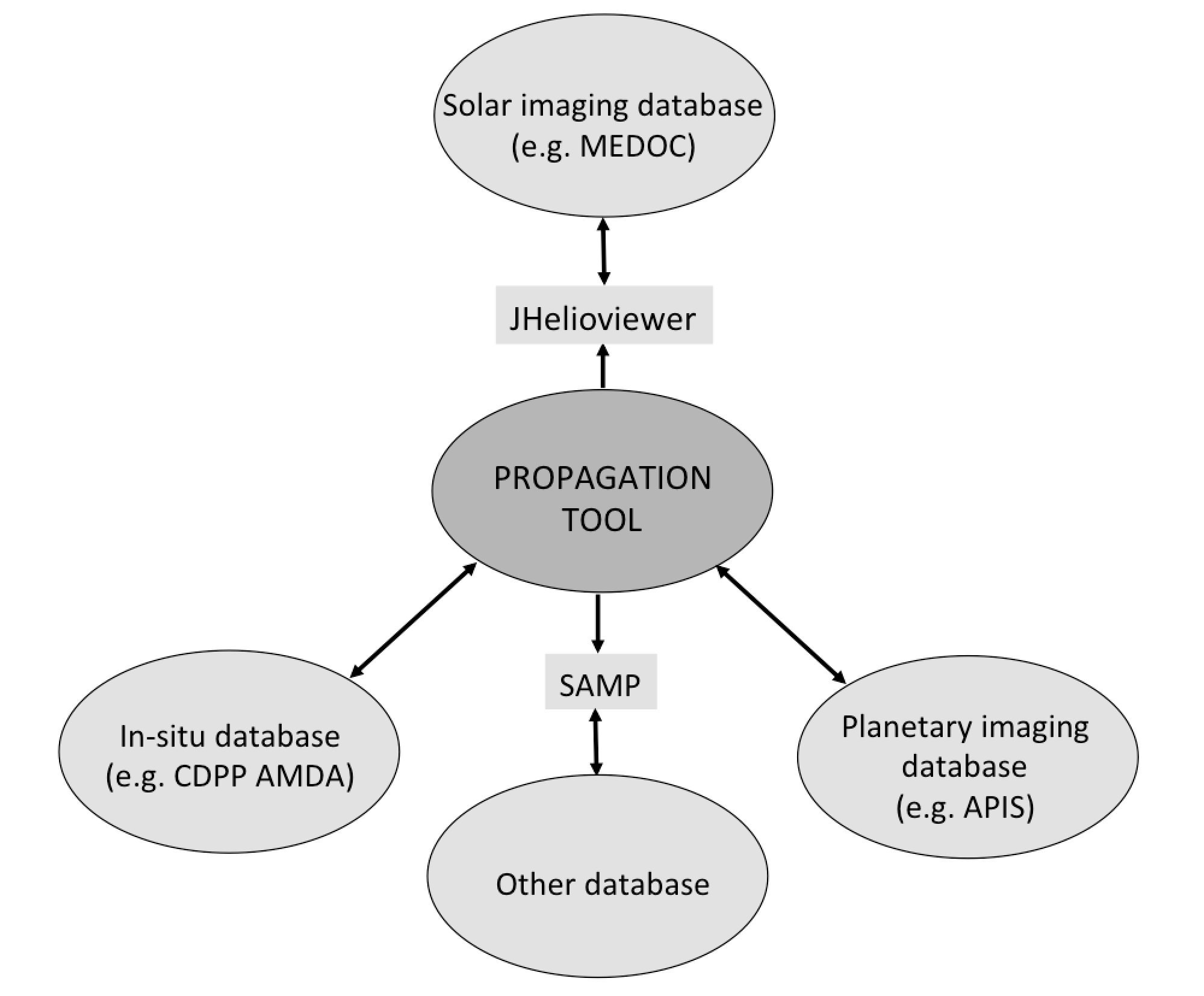}   
 \caption{The links currently working between the propagation tool and various databases that hold in-situ data (CDPP), solar imagery (MEDOC) and planetary imaging (APIS).}
 \label{fig:database}
\end{figure}

\indent The tool was created to support research in heliophysics by establishing connections between different databases that often hold complementary datasets on the sources, evolution and effects of propagating heliospheric structures. The different databases that are currently connected to the Propagation Tool are shown in Figure \ref{fig:database}. They include the MEDOC solar imagery datacenter (\url{https://idoc.ias.u-psud.fr/MEDOC}), the CDPP in-situ datacenter (\url{http://cdpp.eu}) and the Auroral Planetary Imaging and Spectroscopy (APIS) database (\url{http://apis.obspm.fr/}). In addition the tool has access to the Simple Applications Messaging Protocol (SAMP) that allows the propagation tool to exchange data with other tools or users through a central hub and the Virtual Observatory for Planetary Science (VESPA) which can be interrogated to access a number of other databases. Once the trajectory of a heliospheric structure is determined by the tool, the various databases can be interogated to determine if interesting data is available along its trajectory. \\ 

\paragraph{Structure and aim of the paper} After a brief introduction of the different types of propagation that can be carried out with the tool and the corresponding exploitation of heliospheric imaging (section \ref{sec:overview}), we describe the general layout of the components visible by default in the main interface (section \ref{sec:maininterface}). Each component is then described separately by illustrating some of its functionalities (section \ref{sec:paraminterface}) to \ref{sec:Vplotinterface}). We also show how different databases can be interogated from the propagation tool (section \ref{sec:accessdatabases}) and how the tool has been used so far for various studies published in the litterature (section \ref{sec:research}). The aim of the paper is to highlight the rich set of functionalities provided by the tool, a good understanding of how the tool works and its full potential can only be achieved by using it. 

\section{Components of the propagation tool}
\subsection{An overview of the propagation methods:}
\label{sec:overview}
\indent In order to establish connections between the effects of a particular solar wind structure (CME, CIR, streamer blob, jet,...) at different locations in the corona/heliosphere system, the propagation time between the start and end points, as well as the spatial extent of the solar wind structure, must be computed as accurately as possible. The propagation time is calculated once the kinematic and spatial properties of that propagating structure are either specified by (1) the users manually, (2) selection from a catalogue of pre-identified heliospheric structures or (3) determination through a bespoke analysis undertaken with the tool.  Heliospheric imagery, presented in the tool in the form of time-elongation maps (Sheeley et al. 1997; Davies et al. 2009) of solar wind density variations, provides the crucial information needed to confirm the trajectory of the outflowing structure. The tool provides a visualisation interface to these time-elongation maps (often called J-maps). The J-maps extend in elongation from the solar corona, imaged by coronagraphs, through the inner heliopshere, out to 1AU and beyond imaged by heliospheric imagers. The generation of these maps will be discussed in more detail in section \ref{sec:Jmapinterface}. As the signal within a pixel integates along the line-of-sight, a specific elongation does not correspond to a unique height above the solar surface and analysis techniques must be applied to determine the 3-D trajectory of the heliospheric structure of interest. These techniques are usually based on ballistic propagation techniques  (see section \ref{sec:Jmapinterface}).  \\

\indent The Propagation Tool offers three simple types of propagations that allow users to:

\begin{itemize}
    \item     propagate radially-outflowing structures (CMEs) sunward or anti-sunward (Radial Propagation method),
    \item     propagate corotating structures (CIRs) forwards in time (Corotation method),
    \item     propagate solar energetic particles (SEPs) along magnetic fields lines sunward or anti-sunward (SEP propagation method).
\end{itemize}
For the radial propagation and corotation methods, the tool offers three different ways to use J-maps that are extremely useful to evaluate the accuracy of a particular propagation technique by comparing the model with real observations. \\

\begin{figure}[ht!]
 \centering
 \includegraphics[width=0.8\textwidth,clip,angle=0]{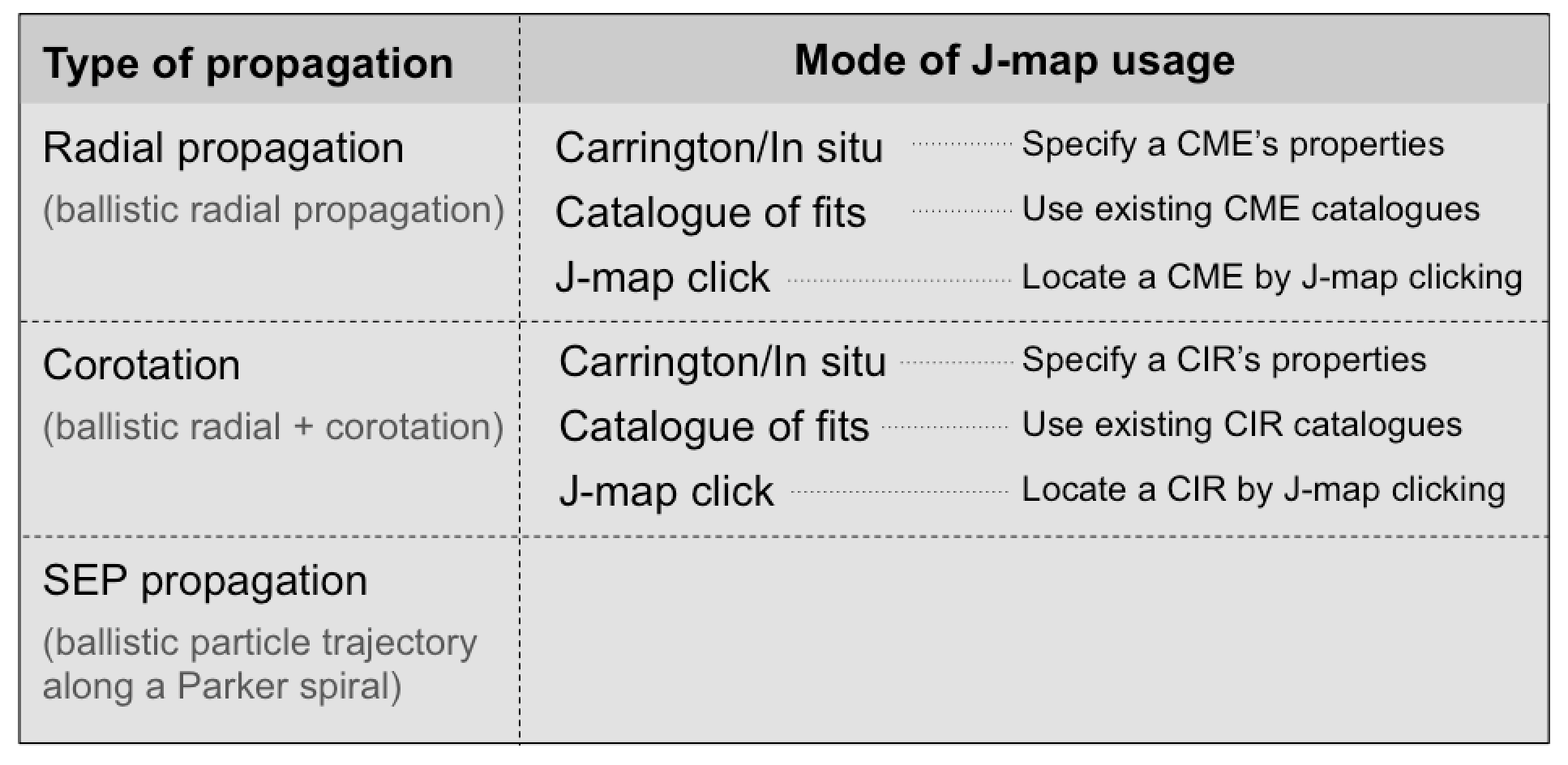}   
 \caption{The three types of propagation and the different ways in which J-maps can be used to track heliospheric structures. Only the radial propagation and the corotation methods allow usage of J-maps. }
 \label{fig:proptech}
\end{figure}

A summary of the different types of propagations and modes of J-map usage are given in Figure \ref{fig:proptech}. We provide examples of the different ways J-maps can be used in section \ref{sec:Jmapinterface}.

For the three propagation methods, we employ the most simple ballistic techniques so that navigation back and forth between imagery and in-situ data browsing remains intuitive. 

\begin{figure}[ht!]
 \centering
 \includegraphics[width=1\textwidth,clip,angle=0]{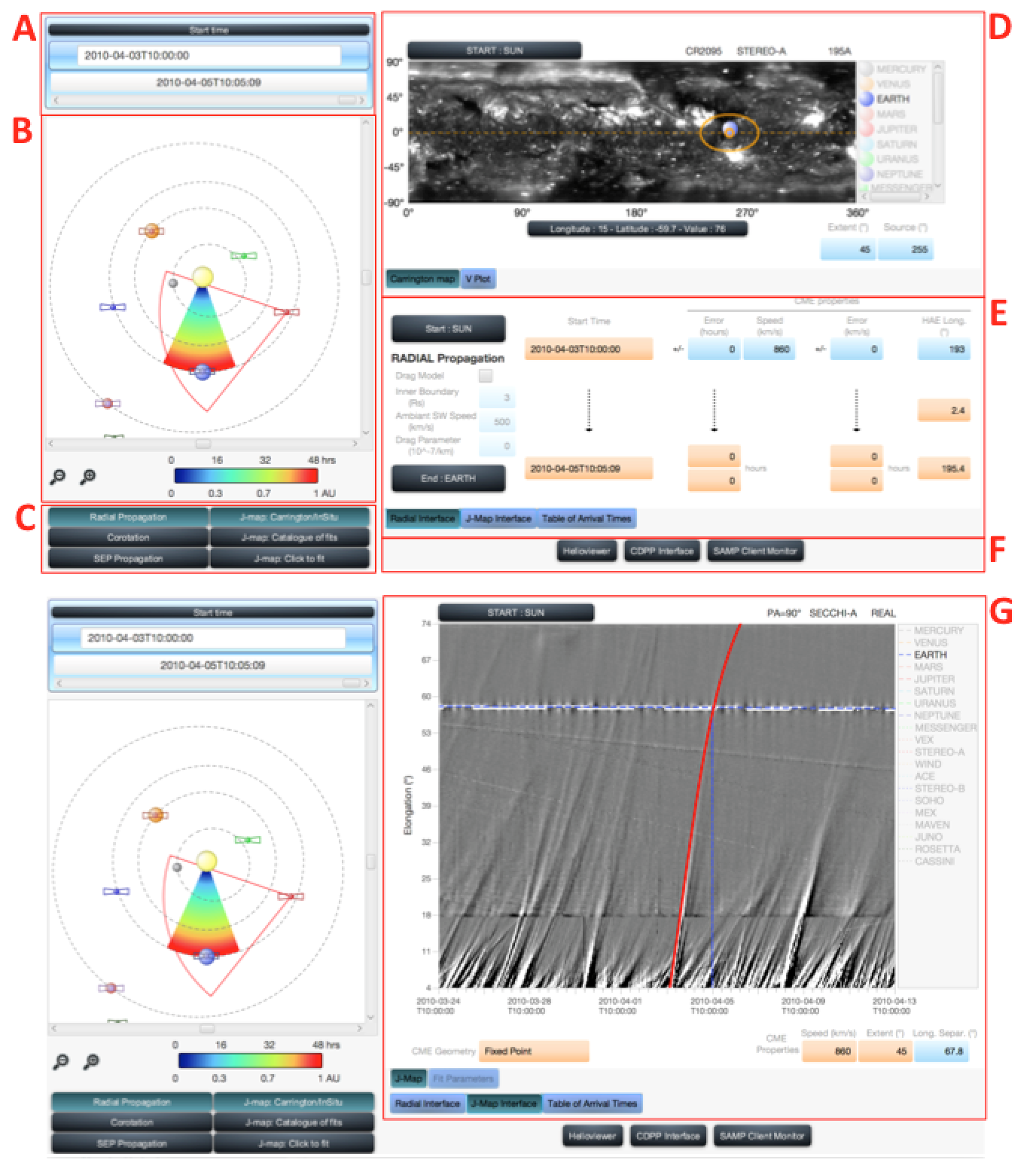}   
 \caption{The basic interface of the propagation tool with seven of its components shown: (B) the ecliptic plane, (C) the propagation type selector, (D) the Carrington map/V plot, (E) the parameter interface, (F) the database selector, (G) the J-map interface.}
 \label{fig:maininterface}
\end{figure}

\section{Components of the tool:}

\subsection{The default interface}
\label{sec:maininterface}
The main interface is shown in Figure \ref{fig:maininterface} and displays seven components of the tool. The set-up of the displayed interface is for the radial propagation method of a radially-outflowing structure such as a CME. For a user-defined time (area A), the orbital positions of various probes, planets and comets are derived from their SPICE kernels and shown in the ecliptic plane (area B). Area E is the interface through which the users defines the starting point of the transient, its speed at that point and its direction of propagation. In the example shown in Figure \ref{fig:maininterface}, the transient is assumed to move outward at a constant speed from Sun to Earth. With the parameters entered in area E, the propagation of the transient is computed by the tool and the result is displayed on the ecliptic plane and if an End point is also entered in area E (here the Earth), the arrival time is computed at that End point. By default, the simplest geometry is assumed for the radially-outflowing transient (area C), it is an arc of a Sun-centered circle moving outward over time. The color-code corresponds to the time, in hours, for the leading edge to reach a certain distance away from the Sun (see color bar). The Earth corresponds to the blue filled circle always situated below the Sun. The map appearing in the top right-hand side (area D) is a Carrington map; it shows a gray-scale map of extreme ultraviolet emission (195 \AA) of the corona from the imager on STEREO-A. The choice of the Carrington rotation number is defined by the launch time of the radially-outflowing transient computed automatically by the tool once the parameters are defined in area E. The extent of the transient projected onto this map is marked by an ellipse which corresponds to the projection of the angular cone sustained by the transient's latitudinal and longitudinal extent. Access to the different databases is available in the bottom right-hand part of the interface (area F). By default, the tool also loads a J-map built with STEREO-A observations. This J-map interface, shown in area G,  replaces areas E and F when the button  'J-map interface' is pressed in area E. The red curve superposed on the J-map tracks the angular position of the center of the radially outflowing CME seen in the ecliptic plane as viewed from STEREO-A. This will be explained in more detail in section \ref{sec:overview}. The field of view displayed on the ecliptic plane corresponds to the combined fields of view of the imagers on STEREO-A that were used to construct the J-map shown in area G. \\

Figure \ref{fig:maininterface} gives the general layout of the interconnected components of the tool that allow, in this example, a simultaneous visualisation of a transient's radial outflow through the inner heliosphere. To summarise, for a given propagation technique, here `Radial propagation' combined with `J-map: Carrington/In situ' selected in area C, a transient's launch time, kinematic properties, extent and direction of propagation can be defined in areas A and E. From these properties, the derived source region and extent of the transient in the low corona are projected on a Carrington map (area D) and its propagation in the interplanetary medium can be visualised on the ecliptic plane (area B) and on a J-map (area G). Once a propagation has been carried out, the tool can interrogate different databases at the relevant times by selecting the buttons in area F. As we shall see, if  `J-map: Catalogue of fits'  or  `J-map: Click to fit' are selected instead in area C, the user defines the kinematic properties and the trajectory of the propagating heliospheric structure (CIR, CME) using the J-map interface and therefore these properties can no longer be entered manually in area E. In the next sections we illustrate the rich set of functions available through the individual components of the tool by considering each component individually. 

\subsection{The parameter interface}
\label{sec:paraminterface}
\begin{figure}[ht!]
 \centering
 \includegraphics[width=0.9\textwidth,clip,angle=0]{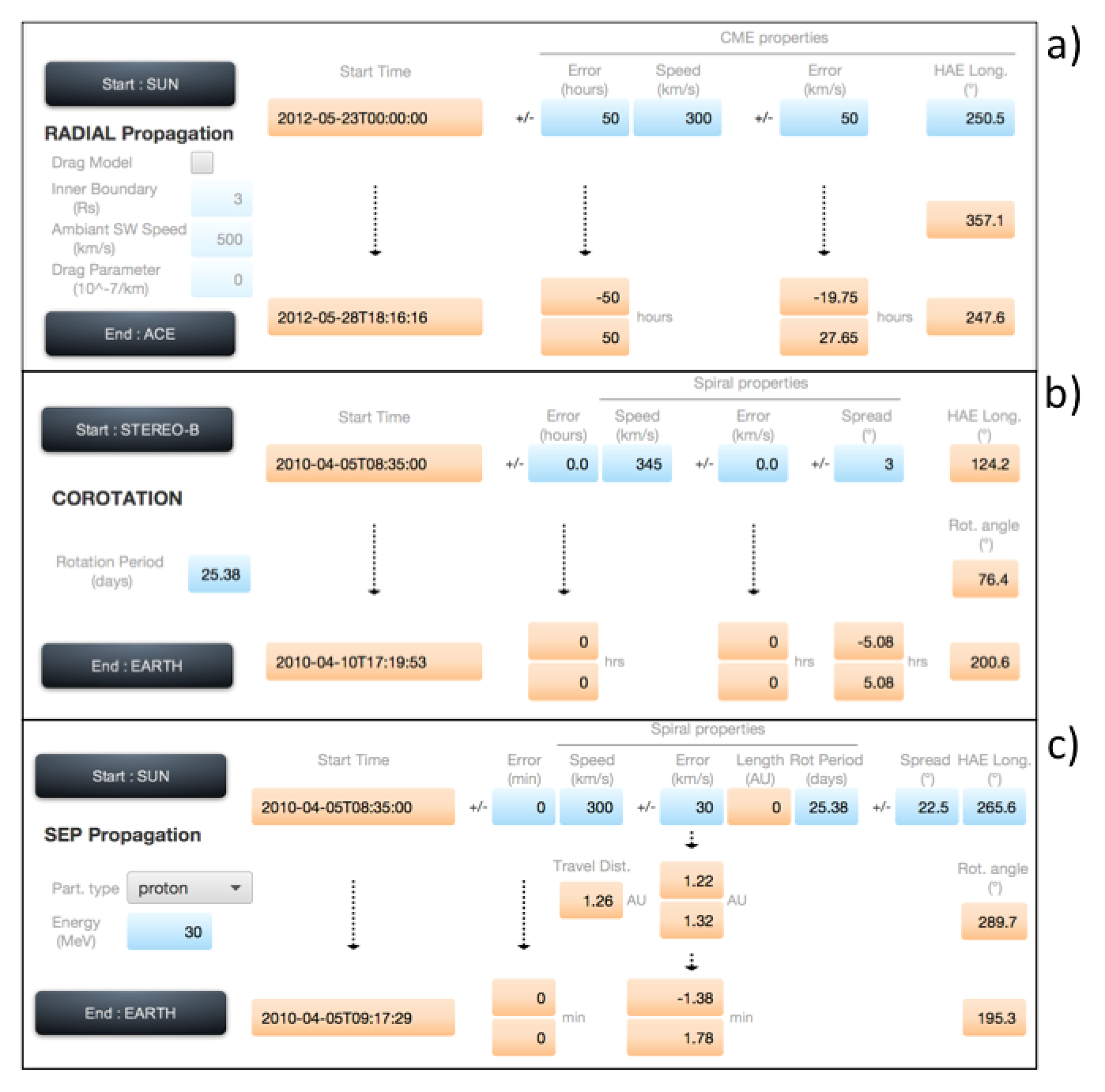}   
 \caption{The three formats of the main interface that allows users to enter the properties of the heliospheric structures correspond to the radial (a), corotation (b) and SEP propagation (c).}
 \label{fig:paraminterface}
\end{figure}

In order to define the start and end points as well as the properties of the CME, CIR or SEP event of interest, the user can enter parameters in area E of Figure \ref{fig:maininterface}. The three different formats of that interface, corresponding to the three propagation methods, are shown in Figure \ref{fig:paraminterface}. Although adapted to the different propagation methods, they all share a similar layout. They correspond to the interface displayed in the 'J-map: Carrington/In situ' mode which allows the parameters of the outflowing transients to be entered manually and not defined by the result of an analysis based on fitting tracks in the J-map ('J-map Catalogue of fits' and 'J-map: Click to fit'). The start and end points of the propagation are input by selecting the corresponding button of the interface, which opens a scroll down menu with a list of selectable planets, probes and the Sun. Those fields of the interface into which values can be entered manually are coloured blue fields. Fields that are not modifiable - because they display either results of the propagation (END row), parameters from pre-defined catalogues (`J-map: Catalogues of fits'), or the results of J-map analysis (`J-map: Click to fit')- are coloured orange. The solar source location and radial speed of the CME (Figure \ref{fig:paraminterface}a), of the CIR (Figure \ref{fig:paraminterface}b) and of the magnetic spiral along which the SEPs propagate (Figure \ref{fig:paraminterface}c) are only selectable in the mode 'J-map Carrington/In situ'. This latter mode offers full flexibility to define the properties of the propagating structures and SEPs. In addition, the radial propagation method (Figure \ref{fig:paraminterface}a) provides an option to activate the drag-based model developped originally by Vrnsak et al.. The drag coefficient and the height in the atmosphere where the user deems necessary to invoke the drag-based model must then be defined. This is the only option available in the tool to vary the speed of a propagated structure. The SEP propagation mode requires  specification of the type and energy of particles to be propagated (Figure \ref{fig:paraminterface}c). Errors in the arrival times are also computed. They account for the uncertainty in the radial speed of the CME and CIRs or the length of the spiral channeling SEPs. Each time a parameter is changed, the propagation is re-run and all components of the tool are updated. The parameters necessary for a propagation to be carried out are:

\begin{itemize}
    \item     Radial propagation: launch time (UT), radial speed (km/s) and source longitude of the CME (in Carrington or HAE coordinates), 
    \item     Corotation: launch time (UT), radial speed (km/s), rotation period of the Sun (in days), source longitude of the CIR (in Carrington or HAE coordinates),
    \item     SEP propagation: launch time (UT), energy (MeV) and source longitude (in Carrington or HAE coordinates) of the SEPs as well as the properties of the Parker spiral defined by the speed of the solar wind (km/s) and the rotation period of the Sun (in days). The propagation time of the particles along these spirals are computed by dividing the speed of the particle to the distance travelled along the Parker spiral between the start and end points. The energy of the particles are of course converted to speed by accounting for relativistic effects.
\end{itemize}

\subsection{The ecliptic plane}
\label{sec:eclipticinterface}

\begin{figure}[ht!]
 \centering
 \includegraphics[width=0.6\textwidth,clip,angle=0]{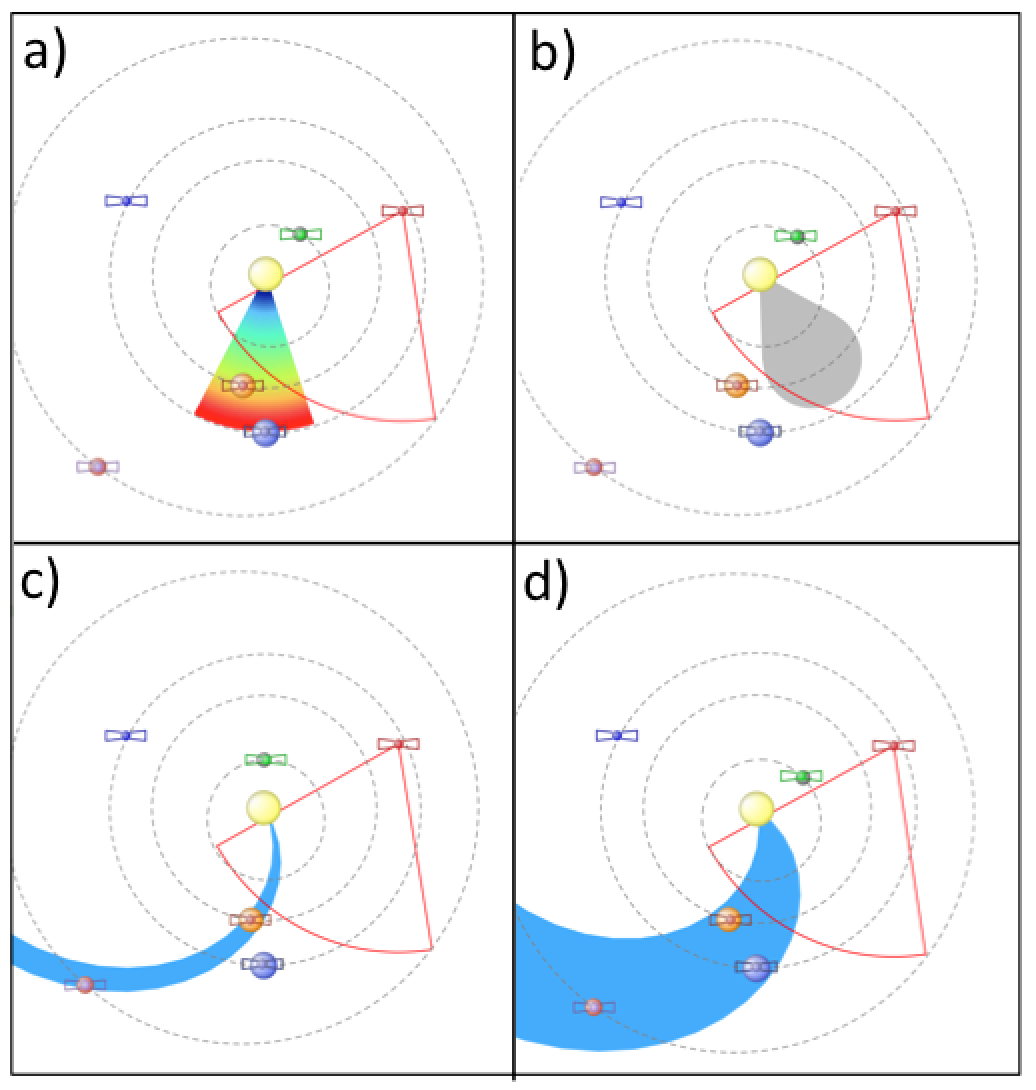}   
 \caption{Views of the ecliptic plane from solar as seen from above on 13 May 2012 at 01:10UT showing the various geometries assumed for the different propagation methods. In all four views, we show the position of planets (circular symbols) and probes (butterflies) as well as the field of view of the combined heliospheric imagers on STEREO-A. The radially propagating structures propagations are tracked as either an  arc centered at the Sun (a) or modelled as an ice-cream cone (b), the corotating structure is assumed to be a corotating spiral (c) and the energetic particle propagation is a fixed spiral centered at the particle beam source and with an angular extent defined by the longitudinal extent of the particle accelerator.}
 \label{fig:ecliptic}
\end{figure}

The interface labelled B in Figure \ref{fig:maininterface} is repeated in Figure \ref{fig:ecliptic} for the three different types of propagation. This figure shows the geometries used to study radially-outflowing transients (panels a,b), corotating structures (c) and beams of energetic particles (d). The horizontal scroll bar at the top of the ecliptic plane can be moved to let time flow, fowards or backwards, and enable the location of the modeled heliospheric structure or SEP stream to be visualised in the ecliptic plane as a function of time.  \\

\textbf{The radial propagation} method exploits two techniques based on different geometries. The simplest geometry assumes that the CME is the arc of a circle with the Sun at its center. As the arc moves radially outwards, with the same speed along its entire front, it forms the color-coded sector visible in Figure \ref{fig:ecliptic}a. The second geometry assumes the CME is the intersection of the 3-D ice-cream cone model with the ecliptic plane, the ice-cream cone model is used routinely to model CMEs in coronagraphic observations. The resulting 2-D shape for the case is shown as the gray-shaded area in Figure \ref{fig:ecliptic}b. Such a shape has been used to located CMEs in coronagraphic in heliospheric images. It has been called the self-similar expansion technique when used to located CMEs in J-maps. Due to the effect of the line of sight integration, the angular distance away from the Sun-observer line of these two CME shapes will be quite different (even for the same speed, direction of propagation and longitudinal extent of the CME), this is discussed in section \ref{sec:Jmapinterface}. The propagation times computed with the two techniques will be different when the end point is situated off the central axis of the outflowing transient because the curvatures of the two arcs are different. The tool accounts for these differences and also considers the orbital motion of the target planet/spacecraft during the propagation time of the CME from the start to the end point.\\
\textbf{The corotation} method assumes that the structure is a corotating Archimedean spiral defined by the speed of the solar wind and the solar rotation period (Figure \ref{fig:ecliptic}c). The computed propagation time between the start and end points folds in the necessary radial motion and corotation. It also accounts for the orbital motion of probes and planets between the start and end times. The latter is particularly important for the propagation of CIRs over many days since the planets and probes can move several degrees in longitude along their orbits over these long time intervals.\\
\textbf{The SEP propagation} assumes that the spatial extent of the beam of energetic particles is represented by a bundle of Parker spirals (Figure \ref{fig:ecliptic}d). The longitudinal extent of these spirals are defined by the user and should correspond to the extent of the particle source region, for instance of a triangulated shock in the low corona (e.g. Rouillard et al. 2016). Solar energetic particles of interest are usually so fast that typically the spiral does not move greatly between the start and end times. Even the low-energy suprathermal electrons with energies of 300eV take at least 3-4 hours to reach Earth along a spiral propagating outwards with a speed of 300 $km/s$. Nevertheless even during that transit time, the spiral will corotate by 2-3 degrees; the tool will tell check that the Earth has not moved out of the corotating Parker spiral between the release and impact times of the particles.

\subsection{The Carrington map}
\label{sec:Carrinterface}
\begin{figure}[ht!]
 \centering
 \includegraphics[width=1\textwidth,clip,angle=0]{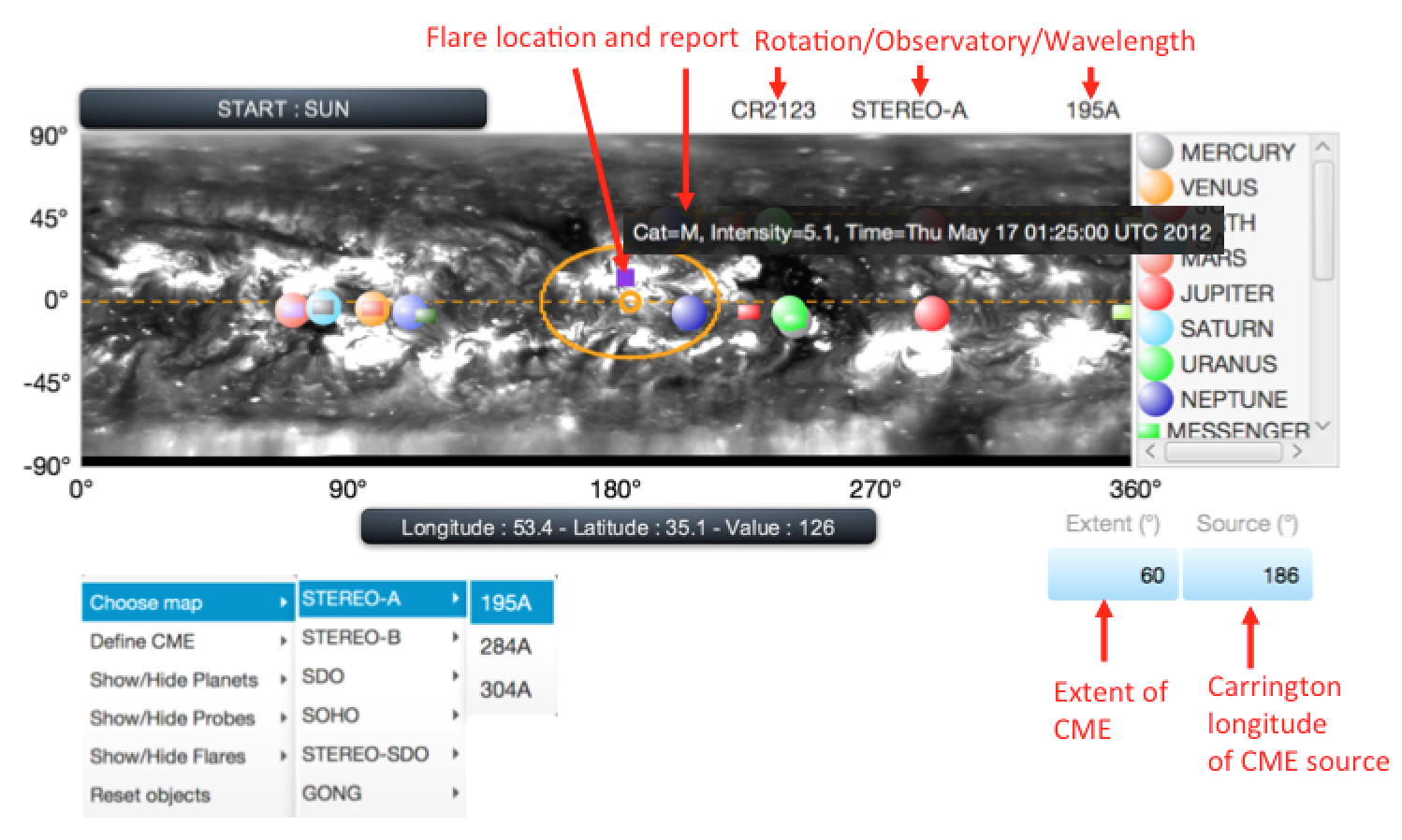}   
 \caption{The Carrington map interface of the propagation tool for, in this case, the radial propagation method. The Carrington map displayed in the background is derived from EUV observations made by the STEREO-A at 195 Angstroom during Carrington Rotation number 2123. The Carrington longitude and heliographic latitude of planets and probes at the launch time of the CME are shown as circles and rectangles, respectively, and defined the legend on the right-hand side. A M-class flare occured near the launch time of the CME and the flare report can be read by placing the mouse cursor on the flare icon (purple square). The menu accessible by a right click on the Carrington map is shown separately below the Carrington map.}
 \label{fig:Carrington_map}
\end{figure}

The interface labelled D in Figure \ref{fig:maininterface} is repeated in Figure \ref{fig:Carrington_map}. It offers a visualisation of Carrington maps of the solar surface in order to ease the definition/identification of the source region of propagating heliospheric structures. The identifiable coronal features are typically coronal holes, active regions and flares. The accessible database of Carrington maps includes maps constructed from extreme ultraviolet and white-light images of the corona as well as maps of magnetic measurements at the photospheric level. In the near future we will also provide results of numerical simulations of solar wind simulations at different altitudes in the atmosphere in order to locate likely source regions of the fast and slow solar winds. A right-click on the Carrington map opens up a menu which is adapted to the propagation method. This menu is shown in Figure \ref{fig:Carrington_map} below the Carrington map for the radial propagation method. This menu provides access to the different types of Carrington maps that can be selected, as well as the the positions of C, M and X-class flares on the solar surface near the estimated launch time of the radially-outflowing transient. This can be used to confirm the release time of a CME for instance or re-define the location of the CME source region by right-clicking on the Carrington map (in the `J-map Carrington/In-situ mode' only). The Carrington coordinates of the planets (circles) and probes at the launch time of the CME can also be projected on the map for the purpose of comparing the longitudes of the CME and the planets/probes. This projection will change when the user uses the time scroll bar in the interface of the ecliptic plane. For the other propagation methods, the projection method is different; in the corotation tool the source location is defined by the longitude of the footpoint location of the CIR spiral at the impact times of the CIR at the different planets/probes. For a typical synodic rotation rate, all planets appear on the Carrington map at roughly the same location with only slight relative displacement corresponding the orbital motion of the impacted objects. For the SEP propagation, the projected position on the Carrington map corresponds to the footpoint location of the spiral at the launch time of the SEPs so that magnetic connectivity of each planet and probe to the low corona is evaluated at the single release time of the particles. Currently the magnetic connectivity assumes a Parker spiral all the way down to the solar surface. We plan, in the near future, to trace magnetic connectivity with more realistic models of the coronal and interplanetary medium. The extent in latitude and longitude of the CME, CIR spiral and the beam of SEPs is shown in an identical way on the Carrington map: they correspond to the same orange ellipse representing the intersection of the geometrical shapes assumed for the different heliospheric structures (CME, CIR spiral and the Parker spirals channeling the SEPs). 

\subsection{V plot:}
\label{sec:Vplotinterface}
\begin{figure}[ht!]
 \centering
 \includegraphics[width=0.9\textwidth,clip,angle=0]{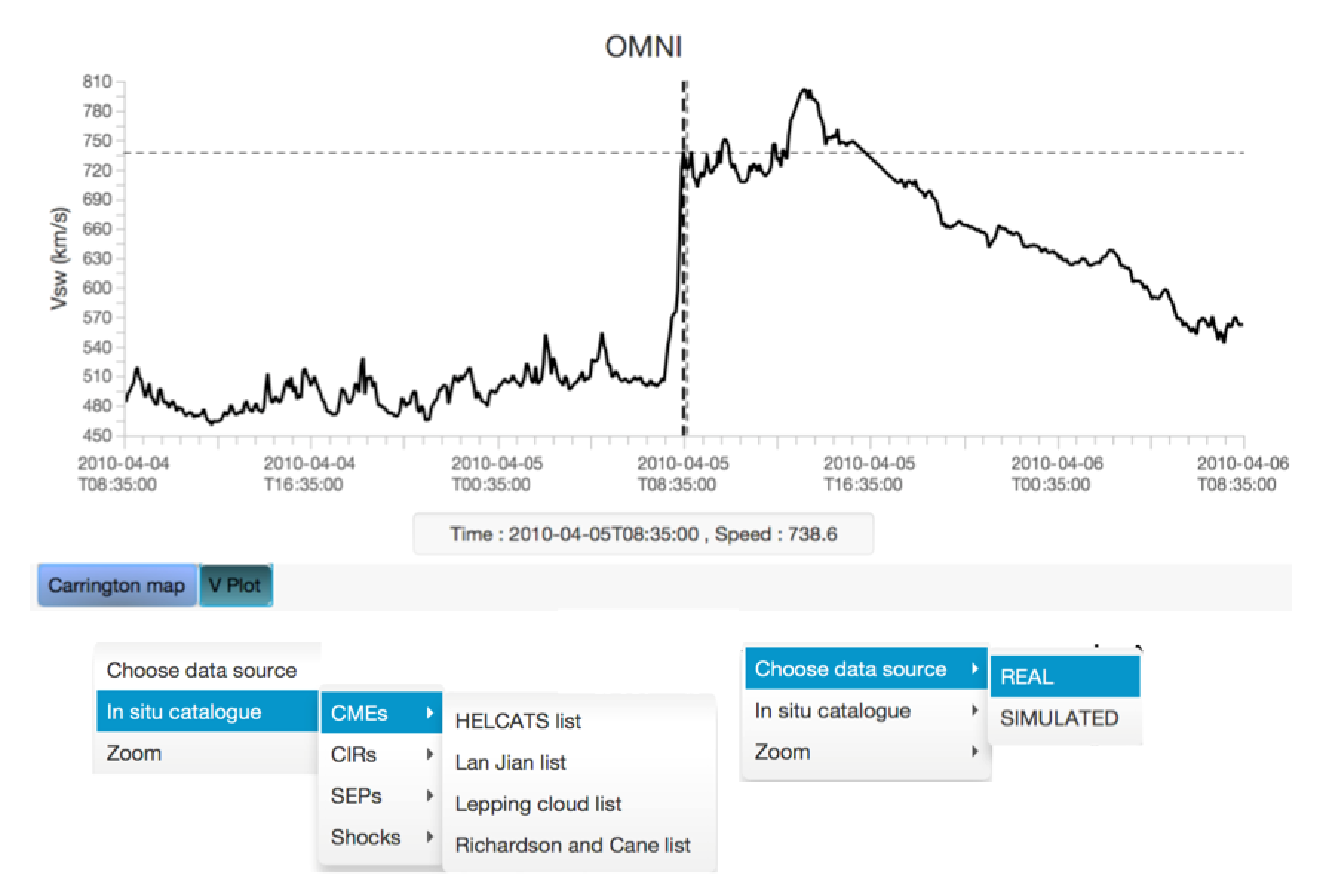}   
 \caption{The Vplot interface accessible in the tool from the Carrington/In situ mode and when the user defines a start point where solar wind speed has either been measured or simulated. The different menus accessible by a right click on the Carrington map is shown separately below the Vplot.}
 \label{fig:Vplot}
\end{figure}

This component, shown in Figure \ref{fig:Vplot} is only accessible in the `J-map: Carrington-In situ' mode. The V-plot replaces the Carrington map when the user selects 'V-plot' and decides to start the propagation from a probe or planet where the radial speed of the heliospheric plasma has been directly measured in-situ or simulated numerically. This functionality allows users to define the propagation speed of the heliospheric structure (CME, CIR, Parker spiral) with direct measurements instead of imagery. The passage of a CME shock over Earth (in OMNI data) is shown in Figure \ref{fig:Vplot}  and corresponds to the CME that emerged from the solar surface on 2010 April 3 (Rouillard et al. 2011a). Once a speed has been selected on the V-plot, all components of the tool are automatically updated.

A cursor allows the user to specify the radial speed of the heliospheric structure propagated with the tool. Scroll down menus are accessible via this interface to select whether real or simulated data should be plotted in the interface.   In addition this component offers the possibility to access a large number of ICMEs, CIR and SEP catalogues, including those  produced by the FP7 HELCATS project.

\begin{figure}[ht!]
 \centering
 \includegraphics[width=0.75\textwidth,clip,angle=0]{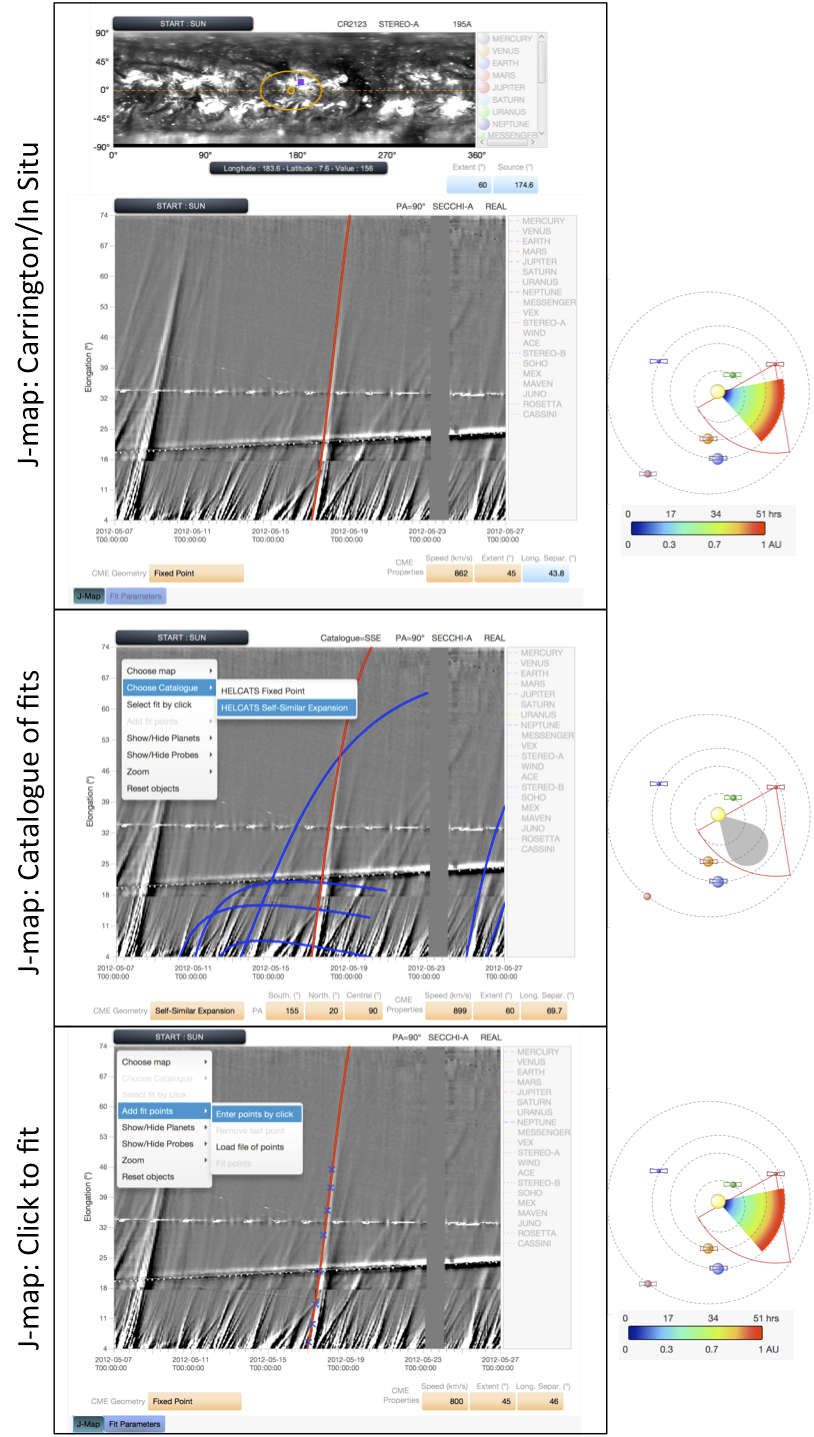}   
 \caption{The three ways in which the J-map interface can be used for the radial propagation of a CME. The top panel also shows the Carrington map used to define the source region of the CME. The direction of propagation is shown in ecliptic plane. }
 \label{fig:Jmap_interface_CME}
\end{figure}

\subsection{The J-map interface:}
\label{sec:Jmapinterface}
The J-map interface is probably the most interesting asset of the propagation tool since it offers an interactive visualisation and exploitation of maps of solar wind outflows obtained from the STEREO spacecraft. The way J-maps are produced with HI-1/2 images is described in e.g. Davies et al. (2009). A simple visualisation of the maps cannot reveal along which heliocentric longitude a particular heliospheric structure is propagating. Indeed each image pixel used to construct a J-map results from the integration of the light scattered by solar wind electrons along the the line of sight defined by that pixel's pointing direction. However the rate of change of the CME's elongation (angular distance to the Sun) with respect to time is to a large extent defined by the longitudinal separation between the CME and the observer (e.g. Rouillard et al. 2008). Using simple geometrical shapes representing a CME, the observed time-varying elongation angle can be fitted to derive a CME's direction of propagation relative to the Sun-Observer line (Rouillard et al. 2008, Lugaz et al. 2010, Davies et al. 2012). The interface allows superposition on the J-maps of the theoretical trajectories obtained from either (1) CME properties (direction of propagation, speed, extent) defined manually by the user (`J-map: Carrington-In situ' mode), (2) from pre-defined catalogues of CME trajectories (`J-map: Catalogue of Fits' mode) or (3) directly fitted by the user (`J-map: Click to Fit' mode). These three modes are selected in the area C of the main interface (Figure \ref{fig:maininterface}). 
The two techniques exploited in the propagation tool to map a CME's heliocentric position into an angular position are the so-called fixed-phi techniques (Sheeley et al. 1999; Rouillard et al. 2008) and the self-similar expansion technique (Davies et al. 2012, Moestl and Davies 2013). The fixed-phi approach assumes that the CME is a single point (hence sometimes also called 'fixed-point technique') situated at the center of the CME arc shown in panel a of Figure \ref{fig:ecliptic}. The self-similar approach assumes that the outermost position of the CME in the heliospheric images is defined by the intersection of the line passing by the observer and which is tangent to the semi-circular structure shown in panel b of Figure \ref{fig:ecliptic}. \\
We present in Figure \ref{fig:Jmap_interface_CME} the three different ways in which the J-map can be used to visualise CME trajectories. At this stage, only ecliptic J-maps are visualised because the tool was devised to connect in-situ measurements, which up to now, have been primarily taken in the ecliptic plane. Extension to other latitudes and to a fully 3-D geometry is planned in preparation for science support of the Solar Orbiter mission (see section \ref{sec:futuremissions}).  Results of using the fixed-phi technique are shown in the top and bottom J-maps shown in Figure \ref{fig:Jmap_interface_CME} and the result of using the self-simiar expansion technique is shown in the middle J-map. Superposed on the J-map in the top panel is the apparent fixed-phi trajectory (as a red line) of the CME that erupted on 17 May 2012 (Rouillard et al. 2016) and whose properties were defined in this example by selecting manually the source region near the location of the M-class flare identified on the Carrington map also shown in that panel (it is the same rotation as in Figure \ref{fig:Carrington_map}). The modeled and real tracks match well for a radial speed of 862 km/s defined manually in the parameter interface and by comparing the observed with the theoretical tracks on the J-map. This approach confirms the idea that a CME propagated outwards from the flare location for a reasonable speed value. Also shown are the time-varying elongations of the Earth, Mars and Venus during the displayed time interval. The `J-map: Carrington-In situ' mode offers the flexibility to start from a spacecraft taking in-situ measurements (through the V-plot interface) to define the time of passage of the CME at that spacecraft. For such an approach the CME trajectory is defined by the longitude of the spacecraft and the speed selected in-situ via the V-plot interface (Figure \ref{fig:Vplot}). \\
The middle panel displays the `J-map: Catalogue of fits' interface. The user can overlay, onto the J-map, tracks from a pre-defined catalogue produced by the FP7 HELCATS project corresponding to either of the two geometries described above. The catalogues are accessible by right-clicking on the J-map through the menu displayed thereon. In this example, the blue lines correspond to the fitted elongation variations obtained with the self-similar expansion technique. To select a particular track, the user right-clicks on the map and chooses the 'Select Track' function and the selected track becomes red. The tool is then updated assuming the catalogue's CME trajectory and its kinematic properties. This includes of course the ecliptic plane shown on the right-hand sides. The HELCATS catalogues were derived using J-maps generated along the CME central axis. Often this does not correspond to the ecliptic plane. Since the tool plots these theoretical tracks on J-maps constructed along the ecliptic plane, there can be a discrepancy between the theoretical and observed tracks due to this latitudinal difference. The PAs considered to fit the CME trajectories as well as the northernmost and southernmost PAs at which the CME was observed in heliospheric images are given below the J-map for the purpose of comparison. More information on these catalogues are given in Harrison et al. (2016).  In order to gain some insight on how different the results of the fixed-phi and the self-similar technique can be, the user is refered to the paper by Moestl and Davies (2013) or by using the propagation tool.  \\
The bottom panel shows the interface when `J-map: Click to Fit' is selected. In this mode, the user can right click on the J-map and selects the function `Enter Points by Click', to enter points situated on the observed track of interest. At this stage, the tool can only fit the selected points with the Fixed-Phi technique and all of the components of the tool are updated automatically once the trajectory is derived. The fitting technique accounts for the motion of the imaging instrument along its orbit over the duration of the fit. Once a CME trajectory is determined, the arrival times of the CME at any planet and many probes can be computed and is made available the interface 'table of arrival times' (not shown here). These arrival times account for the geometry of the CME and the orbital motion of the spacecraft/planet. For long propagation times a planet or probe can move out of the CME's path, the tool will take that into account as well.\\

\begin{figure}[ht!]
 \centering
 \includegraphics[width=0.75\textwidth,clip,angle=0]{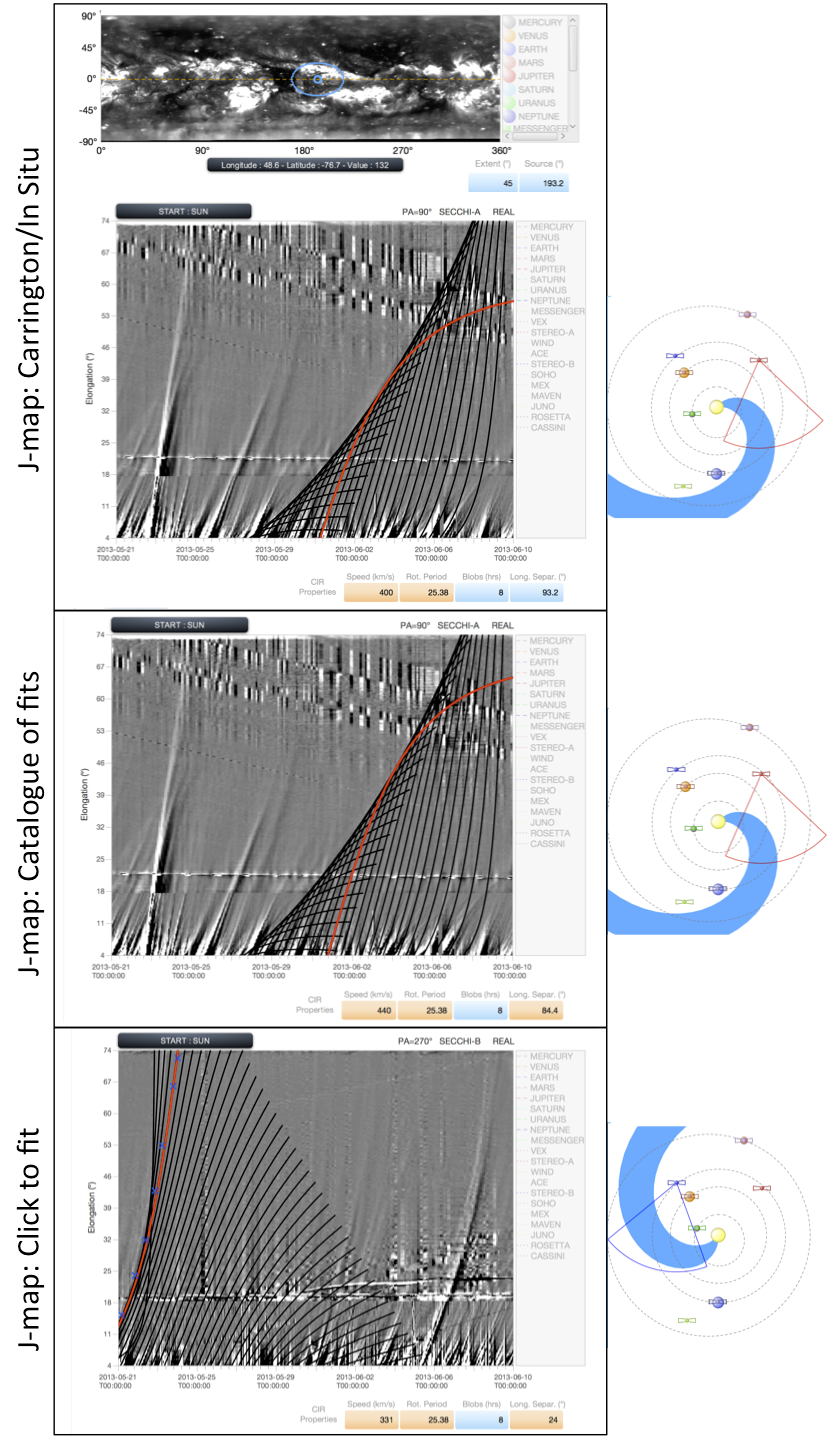}   
 \caption{Same as Figure 8 but for the corotation method.}
 \label{fig:Jmap_interface_CIR}
\end{figure}

The three J-map tools discussed above are also available for propagation by corotation. In J-maps constructed from STEREO-A/STEREO-B HI images, characteristic patterns of converging/diverging tracks appear due to corotating density structures (such as CIRs) as demonstrated by Rouillard et al. (2008, 2010a) and Sheeley et al. (2008b). Each track in any such pattern corresponds to the Thomson-scattered white-light signature of a strong density inhomogeneity (a so-called ‘density blob’) moving radially outward from the Sun. Beyond about 0.5 AU (situated roughly in the overlap region between the HI-1 and HI-2 fields of view), these inhomogeneities become entrained ahead of corotating high-speed streams (e.g. Rouillard et al., 2008). Because these density structures are emitted by a spatially limited source region on the Sun, they rapidly form a spiral of density inhomogeneities in the interplanetary medium. This spiral is analogous to the Parker spiral formed by the interplanetary magnetic field, as both trace approximately the locus of plasma emitted by a single corotating source region on the Sun. The three J-map tools available to analyse these corotating structures are displayed in Figure \ref{fig:Jmap_interface_CIR}. In these examples, they all show the same passage of a CIR that formed due to an extensive low-latitude coronal hole imaged by NASA's Solar Dynamics Observatory between the 25 May 2013 (appearance on the East limb) and the 03 June 2013 (disappearance behind the West limb). The coronal hole was also imaged by the STEREO-A spacecraft and the associated Carrington map (rotation 2137) is shown in the top panel situated above the J-map. This CIR has been analysed extensively by Sanchez-Diaz et al. (2016) because multi-point observations of the corona during this CIR passage revealed, for the first time, the formation height of the density inhomogeneities (blobs) entrained by the CIR. The source region of a CIR can often be located near the Western edge of these low-latitude coronal hole, where we expect the fast and slow solar wind to form at adjacent longitudes and to become radially aligned in response to solar rotation at high altitudes. To reconstruct the CIR pattern, the tool assumes that at a single blob is emitted along a longitude near the Western boundary of the coronal hole (in the ecliptic), this produces the red track on the J-map. The CIR pattern that results from a succession of blobs emitted every 8 hours from that same corotating longitude is shown as the black curves in the top panel. A good match between the family of black curves and the observed pattern of converging tracks confirms the tool's localisation of the CIR. Plotnikov et al. (2016) showed that at solar minimum all CIRs that eventually impact a spacecraft can be located using the heliospheric imagers. Hence, acatalogue of CIRs could be derived for the FP7 HELCATS project and the CIR described above is part of this catalogue (Plotnikov et al. 2016). The catalogue gives an average speed for the modeled CIR of 440 km/s and confirms the source region near the Western boundary of the giant coronal hole already mentioned. The selection of this catalogue is done in an analogous manner to the CME catalogue by right-clicking on the J-map. The bottom panel shows the CIR pattern left by the same CIR a few days early in the STEREO-B imager. CIRs are difficult to identify in STEREO-B images because the pattern of diverging tracks is not always visible. One feature that nearly always appears is however the track left by the CIR when the tangent to the CIR curve passes by STEREO-B (Sheeley and Rouillard 2010). This configuration occurs during an approximately 5-day long time-window. The associated track is fitted in Figure \ref{fig:Jmap_interface_CIR} to the clicked blue crosses. The source longitude of the CIR passing in STEREO-B's imager is confirmed to be situated near the Western boundary of the coronal hole shown in the Carrington map. Once a CIR is located in the J-map, it can be propagated to any planet or probe in the clock-wise direction of solar rotation only. Analogous to the radial propagation, the orbital motion of the spacecraft and probes are accounted for when the tool calculates the impact times. \\
The J-maps cannot directly provide information on SEP propagation, hence the tool does not offer the possibility to access the J-map interfaces when the SEP propagation is selected.

\subsection{Access to databases:}
\label{sec:accessdatabases}
\begin{figure}[ht!]
 \centering
 \includegraphics[width=0.85\textwidth,clip,angle=0]{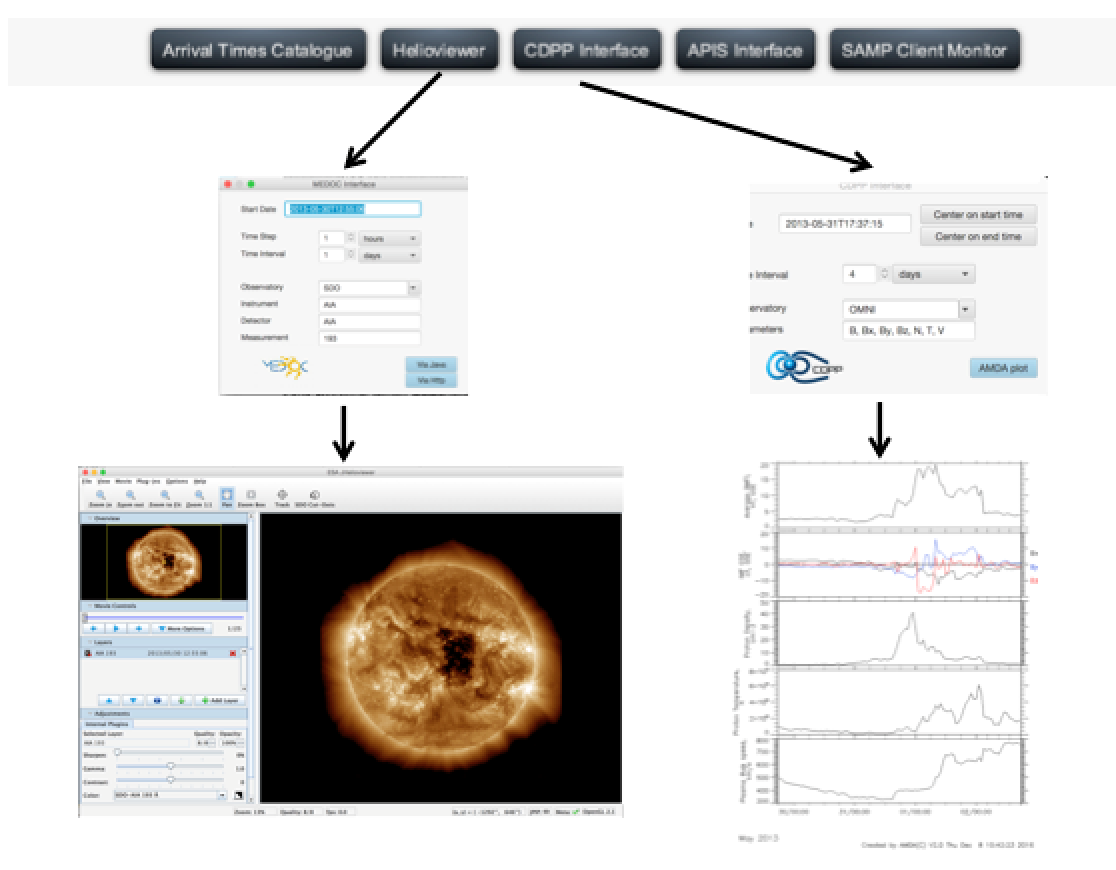}   
 \caption{The steps followed in the tool to launch JHelioviewer and the AMDA tool webservice to visualise remote-sensing observations stored at MEDOC and the in-situ measurements stored at the CDPP, respectively. The databases are accessed by clicking on the black buttons in area F on the main interface (Figure 3). }
 \label{fig:Databases2}
\end{figure}

As mentioned in the introduction, the propagation tool was created to provide a simple way to propagate heliospheric structures in order to connect remote-sensing observations with in-situ measurements. Once a propagation has been carried out with the tool and a launch time has been estimated at the Sun, it is possible to launch two versions of the ESA/NASA Helioviewer interface straight from the propagation tool; these are the Helioviewer webpage and the java webbased interface called JHelioviewer. Both versions were modified to read data stored at the MEDOC datacenter with the aim of providing a back-up site in case data becomes unavailable at the two other different servers running Helioviewer (Goddard Space Flight Center, Royal Observatory Belgium). JHelioviewer offers advanced viewing options, including running and base difference images, a side-by-side (multiview) mode, image projections such as orthographic, latitudinal and polar, timeline displays of 1D and 2D datasets, synchronized with image time series. In addition to compare the launch times of the heliospheric structures propagated in the propagation tool, JHelioviewer can display features and events from the Heliophysics Events Knowledgebase, and alerts from the COMESEP system. The steps followed to launch JHelioviewer are shown in Figure \ref{fig:Databases2}. They begin by selecting the black button labelled 'Helioviewer'. This opens a pop-up window where the user enters a time of interest, a spacecraft, an instrument and the desired type of observation (EUV, white light...). Depending on whether the http or java version of Helioviewer is then selected, the tool will either launch the webpage or a version of JHelioviewer (preinstalled on the user's computer) with the requested images/movies. \\

To browse in situ data at the impact times predicted by the propagation tool a similar procedure is followed by pressing  on the black button labelled 'CDPP interface'. In the launched pop-up, the times, spacecraft and the types of datasets are selected manually. This generates a requests via webservice to the AMDA tool at CDPP, AMDA then creates a plots that is displayed as a page in the user's web browser. The order of solar wind parameters displayed in the created plot are pre-defined. To carry out a more in depth analysis of the in-situ data, the user can launch the full data mining environment offered by the AMDA tool at CDPP (\url{amda.cdpp.eu}). AMDA is a multidataset visualisation and download software, it can do automated event search and data mining, it provides access to existing catalogues and allows users to produce their own catalogues. Most of the OMNIweb database can be accessed via AMDA.\\
 
The interoperability of the Propagation Tool has been recently expanded with a direct connection to the APIS database (http://apis.obspm.fr/) of auroral images. The connection is made possible by a data exchange protocole called EPN-TAP, developed and promoted by the VESPA infrastructure of Europlanet 2020 RI (\url{http://www.europlanet-vespa.eu/}). When a solar wind structure impacts a planet (most appropriately a giant planet) hitting the 'APIS Interface' button launches a search in APIS for the same target and for an interval of time bracketing the structure arrival time. If the search returns results (in the form of raw/processed/projection images) they can be either browsed (as thumbnails are available), or visualized (in case of pdf or png), or selected for full display and further analysis in companion tools like Aladin or Topcat (data are then sent via the IVOA SAMP protocole). These new capabilities will be discussed in a future publication that presents in more detail the recent catalogues that have been produced by international teams and integrated in the propagation tool to further support research in planetary space weather.

\section{Current and future research with the propagation tool:}
\subsection{Recent studies}
\label{sec:research}
This article has presented the core functionalities available in the tool. The large number of links between the numerous components of the interface provide a very rich combination of possible ways in which imagery, modelling and in-situ data can be used to address a number of research topics. One of the major advantages of this tool resides in the speed at which datasets normally accessed via many different websites can be compared in a few clicks only. This is reflected in the variety of different studies that have been recently published that have carried out part of their analysis on results using obtained with the propagation tool. They include studies investigating the long-term variations of corotating density structures (Plotnikov et al. 2016a), the relation between CMEs and solar energetic particles (Salas-Matamoros et al. 2016; Rouillard et al. 2016), the relation between CMEs and solar gamma-ray events (Plotnikov et al. 2016b) as well as the effect of a CME at many planets on its way to outer heliosphere (Witasse et al. 2016).\\
The simple propagation techniques and wide range of connected datasets also make the tool useful for university lecturing and doctoral and post-doctoral summer training schools. The propagation tool was used interactively at the course entitled 'Heliospheric physical processes for understanding Solar-Terrestrial Relations' held in September 2015 in the International School on Space Science in L'Acquila, Italy \url{http://www.cifs-isss.org/pastcourses.asp}).\\

\section{Future datasets:}
\label{sec:futuremissions}
The propagation tool exploits heliospheric imagery to connect imagery with in-situ measurements. The upcoming Solar Probe Plus and Solar Orbiter missions will also carry heliospheric  imagers, they are the WISPR (Vourlidas et al. 2015) and SoloHI (Howard et al. 2010) instruments. We plan to modify the propagation tool to include images of these two instruments. Because the orbit of Solar Orbiter will eventually move out of the ecliptic planes, all the components of the propagation tool will need to be extended to 3-D, this includes tracking techniques, presentation of the propagation in the ecliptic plane and the way we visualise J-maps. We also hope to include ground-based imagery such as spectro-polarimetric measurements obtained by CoMP instrument in Hawaii and soon the CLIMSO instrument at Pic du Midi which provide additional information on the topology of the coronal magnetic field.

\paragraph{Acknowledgments:}
We acknowledge support from the plasma physics data center (Centre de Donnees de la Physique des Plasmas; CDPP; \url{http://cdpp.eu/}), the Virtual Solar Observatory (VSO; \url{http://sdac.virtualsolar.org}), the Multi Experiment Data $\&$ Operation Center (MEDOC; \url{https://idoc.ias.u-psud.fr/MEDOC}), the French space agency (Centre National des Etudes Spatiales; CNES; \url{https://cnes.fr/fr}) and the space weather team in Toulouse (Solar-Terrestrial Observations and Modelling Service; STORMS; \url{https://stormsweb.irap.omp.eu/}). This includes the data mining tools AMDA (\url{http://amda.cdpp.eu/}) and the propagation tool (\url{http://propagationtool.cdpp.eu}). R.F.P., I.P. and ESD acknowledge financial support from the HELCATS project under the FP7 EU contract number 606692. Europlanet 2020 RI has received funding from the European Union's Horizon 2020 research and innovation programme under grant agreement No 654208. The \emph{STEREO} \emph{SECCHI} data are produced by a consortium of \emph{RAL} (UK), \emph{NRL} (USA), \emph{LMSAL} (USA), \emph{GSFC} (USA), \emph{MPS} (Germany), \emph{CSL} (Belgium), \emph{IOTA} (France) and \emph{IAS} (France).\\


\begin{thebibliography}{10}
\expandafter\ifx\csname url\endcsname\relax
  \def\url#1{\texttt{#1}}\fi
\expandafter\ifx\csname urlprefix\endcsname\relax\def\urlprefix{URL }\fi
\expandafter\ifx\csname href\endcsname\relax
  \def\href#1#2{#2} \def\path#1{#1}\fi

\bibitem{Eyles09}
C.~J. {Eyles}, et~al., {The Heliospheric Imagers Onboard the STEREO Mission},
  Sol. Phys. 254 (2009) 387--445.
\newblock \href {http://dx.doi.org/10.1007/s11207-008-9299-0}
  {\path{doi:10.1007/s11207-008-9299-0}}.

\bibitem{Howard08}
R.~A. {Howard}, et~al., {Sun Earth Connection Coronal and Heliospheric
  Investigation (SECCHI)}, Space Sci. Rev. 136 (2008) 67--115.
\newblock \href {http://dx.doi.org/10.1007/s11214-008-9341-4}
  {\path{doi:10.1007/s11214-008-9341-4}}.

\bibitem{Davies09}
J.~A. {Davies}, et~al., {A synoptic view of solar transient evolution in the
  inner heliosphere using the Heliospheric Imagers on STEREO}, Geophys. Res.
  Lett. 36 (2009) L02102.
\newblock \href {http://dx.doi.org/10.1029/2008GL036182}
  {\path{doi:10.1029/2008GL036182}}.

\bibitem{Lugaz10}
N.~{Lugaz}, {Accuracy and Limitations of Fitting and Stereoscopic Methods to
  Determine the Direction of Coronal Mass Ejections from Heliospheric Imagers
  Observations}, Sol. Phys. 267 (2010) 411--429.
\newblock \href {http://arxiv.org/abs/1010.1949} {\path{arXiv:1010.1949}},
  \href {http://dx.doi.org/10.1007/s11207-010-9654-9}
  {\path{doi:10.1007/s11207-010-9654-9}}.

\bibitem{Moestl13}
C.~{Moestl}, J.~A. {Davies}, {Speeds and Arrival Times of Solar Transients
  Approximated by Self-similar Expanding Circular Fronts}, Sol. Phys. 285
  (2013) 411--423.
\newblock \href {http://arxiv.org/abs/1202.1299} {\path{arXiv:1202.1299}},
  \href {http://dx.doi.org/10.1007/s11207-012-9978-8}
  {\path{doi:10.1007/s11207-012-9978-8}}.

\bibitem{Plotnikov16a}
I.~{Plotnikov}, et~al., {Long-Term Tracking of Corotating Density Structures
  Using Heliospheric Imaging}, Solar Physics 291 (2016) 1853--1875.
\newblock \href {http://arxiv.org/abs/1606.01127} {\path{arXiv:1606.01127}},
  \href {http://dx.doi.org/10.1007/s11207-016-0935-9}
  {\path{doi:10.1007/s11207-016-0935-9}}.

\bibitem{Plotnikov16b}
I.~{Plotnikov}, A.~{Rouillard}, G.~{Share}, {Far-side Coronal Mass Ejections:
  shock front magnetic connectivity to the visible disk and the origin of
  long-duration gamma-ray flares}, Astronomy and Astrophysics (2016) submitted.

\bibitem{Rouillard16}
A.~P. {Rouillard}, et~al., {Deriving the Properties of Coronal Pressure Fronts
  in 3D: Application to the 2012 May 17 Ground Level Enhancement}, Astrophys.
  J. 833 (2016) 45.
\newblock \href {http://arxiv.org/abs/1605.05208} {\path{arXiv:1605.05208}},
  \href {http://dx.doi.org/10.3847/1538-4357/833/1/45}
  {\path{doi:10.3847/1538-4357/833/1/45}}.

\bibitem{Rouillard11a}
A.~P. {Rouillard}, et~al., {Interpreting the Properties of Solar Energetic
  Particle Events by Using Combined Imaging and Modeling of Interplanetary
  Shocks}, Astrophys. J. 735 (2011) 7.
\newblock \href {http://dx.doi.org/10.1088/0004-637X/735/1/7}
  {\path{doi:10.1088/0004-637X/735/1/7}}.

\bibitem{Rouillard11b}
A.~P. {Rouillard}, {Relating white light and in situ observations of coronal
  mass ejections: A review}, Journal of Atmospheric and Solar-Terrestrial
  Physics 73 (2011) 1201--1213.
\newblock \href {http://dx.doi.org/10.1016/j.jastp.2010.08.015}
  {\path{doi:10.1016/j.jastp.2010.08.015}}.

\bibitem{Rouillard08}
A.~P. {Rouillard}, et~al., {First imaging of corotating interaction regions
  using the STEREO spacecraft}, Geophys. Res. Lett. 35 (2008) L10110.
\newblock \href {http://dx.doi.org/10.1029/2008GL033767}
  {\path{doi:10.1029/2008GL033767}}.

\bibitem{Salas16}
C.~{Salas-Matamoros}, K.-L. {Klein}, A.~P. {Rouillard}, {Coronal mass
  ejection-related particle acceleration regions during a simple eruptive
  event}, A. and A. 590 (2016) A135.
\newblock \href {http://dx.doi.org/10.1051/0004-6361/201528015}
  {\path{doi:10.1051/0004-6361/201528015}}.

\bibitem{Sanchez16}
E.~{Sanchez-Diaz}, et~al., {Observational Evidence for the Associated Formation
  of Blobs and Raining Inflows in the Solar Corona}, Astrophys. J. Lett. 835
  (2017) L7.
\newblock \href {http://arxiv.org/abs/1612.05487} {\path{arXiv:1612.05487}},
  \href {http://dx.doi.org/10.3847/2041-8213/835/1/L7}
  {\path{doi:10.3847/2041-8213/835/1/L7}}.

\bibitem{Sheeley10}
N.~R. {Sheeley}, Jr., A.~P. {Rouillard}, {Tracking Streamer Blobs into the
  Heliosphere}, Astrophys. J. 715 (2010) 300--309.
\newblock \href {http://arxiv.org/abs/1006.5379} {\path{arXiv:1006.5379}},
  \href {http://dx.doi.org/10.1088/0004-637X/715/1/300}
  {\path{doi:10.1088/0004-637X/715/1/300}}.

\bibitem{Sheeley08}
N.~R. {Sheeley}, Jr., et~al., {SECCHI Observations of the Sun's Garden-Hose
  Density Spiral}, Astrophys. J. Lett. 674 (2008) L109.
\newblock \href {http://dx.doi.org/10.1086/529020} {\path{doi:10.1086/529020}}.

\bibitem{Sheeley97}
N.~R. {Sheeley}, et~al., {Measurements of Flow Speeds in the Corona Between 2
  and 30 Rs}, Astrophys. J. 484 (1997) 472--478.

\bibitem{Vrsnak13}
B.~{Vr{\v s}nak}, et~al., {Propagation of Interplanetary Coronal Mass
  Ejections: The Drag-Based Model}, Sol. Phys. 285 (2013) 295--315.
\newblock \href {http://dx.doi.org/10.1007/s11207-012-0035-4}
  {\path{doi:10.1007/s11207-012-0035-4}}.

\bibitem{Witasse16}
O.~{Witasse}, et~al., {Interplanetary coronal mass ejection observed at
  STEREO-A, Mars, comet 67P/Churyumov-Gerasimenko, Saturn, and New Horizons
  en-route to Pluto. Comparison of its Forbush decreases at 1.4, 3,1 and 9.9
  AU}, Journal of Geophys. Res.  submitted.

\end{thebibliography}
%

\end{document}